%% file: vlasov_pop.tex
\documentclass[aip,showpacs,preprintnumbers,amsmath,amssymb]{revtex4-1}

\usepackage{graphicx}
\usepackage{dcolumn}
\usepackage{bm}

\begin{document}

\preprint{}

\title{A Vlasov Algorithm Derived from Phase Space Conservation}

\author{Jonathan P. Edelen}%
 \author{Stephen D. Webb}%
  \email{swebb@radiasoft.net}
 \homepage{www.radiasoft.net}
\affiliation{RadiaSoft, LLC}
 \altaffiliation[]{}

\date{\today}

\begin{abstract}
Existing approaches to solving the Vlasov equation treat the system as a partial differential equation on a phase space grid, and track in either an Eulerian, Lagrangian, or semi-Lagrangian picture. We present an alternative approach, which treats the Vlasov equation as a conservative flow on phase space, and derives its equations of motion using particle-pushing algorithms akin to particle-in-cell methods. Deposition to the grid is determined from the convolution of local basis functions. This approach has the benefit of allowing flexible definitions in the grid, which are decoupled from how the phase space flow evolves. We present numerical examples and comment on the various properties of the algorithm.
\end{abstract}

\pacs{}
\maketitle

\section{Introduction}
%
\input{sections/introduction.tex}

\section{The Vlasov Equation as Flows on Phase Space}
%
\input{sections/background.tex}

\section{An Algorithm Based on Phase Space Density Conservation}
%
\input{sections/algorithm.tex}

\section{Numerical Examples}
\input{sections/examples.tex}
\input{sections/pendulum.tex}
\input{sections/fel.tex}

\section{Discussion}
%
\input{sections/discussion.tex}

\section{Acknowledgements}

This work was supported in part by the Department of Energy Office of Science, Office of Basic Energy Science under contract no. DE-SC0017161.

\bibliography{vlasov_pop.bib}

\end{document}

%% file: sections/introduction.tex


The core problem of plasma physics is to solve for the evolution of the phase space density under external and self-consistent forces. Where analytic techniques, such as linearized Vlasov, fail, we are required to resort to numerical methods. The two dominant methods for computing the evolution of self-consistent plasma dynamics are the  particle-in-cell approach -- which follows the trajectories of macro-particles that sample the phase space -- and the solution of the Vlasov equation on a numerical grid -- which advects the phase space density on a mesh using various methods. 

Particle-in-cell methods~\cite{birdsall_langdon:85,hockney_eastwood:89} trace macro-particles  that sample the phase space distribution as they move in self-consistent and external fields. This approach is quite mature, and has been used to solve a variety of self-consistent many-body problems ranging from semiconductor transport to large-scale astrophysics. This approach has the benefit of being substantially faster than Vlasov solvers, but are artificially noisy due to the down-sampling of the number of particles. The noise is therefore amplified by a factor $\sim \sqrt{\frac{N_{\textsf{real}}}{N_{\textsf{macro.}}}}$

Vlasov solvers treat phase space density as a fluid, and advect the fluid based on the Vlasov equation. These approaches treat the evolution of phase space by solving the Vlasov partial differential equation. Early approaches, for example, used a local Chebyshev polynomial basis~\cite{shoucri_knorr:1974}, or using splitting and spline or Fourier interpolation~\cite{cheng_knorr:1976}. Finite element approaches have also been used~\cite{zgb1, zgb2}. Modern approaches, such as flux balancing~\cite{filbet_sonnendrucker_bertrand:2001}, use conservation of phase space volume and local numerical integrals to implicitly back-track the characteristic trajectories of a local phase space volume. These characteristic methods~\cite{shoucri:2008} rely on interpolation methods to compute the characteristics, which can be computationally expensive. Although the Vlasov equation is a partial differential equation describing the evolution of a phase space fluid, the evolution of the phase space density is taken not from the fluid equation, but by tracing the characteristics of each volume of phase space, conceptually closer to particle-in-cell methods.

We present here a hybrid method, which evolves the phase space density on a mesh by tracing trajectories of those mesh points using particle-in-cell-like particle pushes. We start by reviewing the formal solution of the Vlasov equation in terms of flows of points in phase space along time. We then translate the flow and phase space conservation into a discrete mesh and make contact with particle-in-cell particle pushers, such as the Boris pusher or symplectic integrator methods. We conclude by demonstrating the method on two examples -- trapped and streaming particles in the pendulum equation, and the high-gain one-dimensional free-electron laser model~\cite{bonifacio_pellegrini_narducci:84}.

%% file: sections/background.tex


The collisionless Vlasov equation can be formulated as a fluid equation on a $2 \times D$-dimensional phase space. The statement of the Vlasov equation as a partial differential equation is
\begin{equation}
\partial_t f + \mathbf{z}' \cdot \nabla_{\mathbf{z}} f = 0
\end{equation}
where $\mathbf{z} = (\mathbf{x}, \mathbf{v})$ are the configuration space coordinates. For Hamiltonian systems with Hamiltonian $\mathcal{H} = p^2/2 + V(x)$, for example, this becomes
\begin{equation}
\partial_t f + p \partial_x f + \left (\partial_x V\right ) \partial_p f = 0.
\end{equation}
If $\mathbf{z}' = \mathbf{v}(\mathbf{z}, t)$, then the solution for the individual trajectories can be given by a flow $\phi$, such that $\mathbf{z}(t) = \phi_{t_0 \rightarrow t} \circ \mathbf{z}(t_0)$, and the Vlasov equation can be solved by the method of characteristics. Thus, the Vlasov equation is equivalent to the conservation of phase space over a flow $\phi$ that maps an initial coordinate $\mathbf{z}^{in.}$ to $\mathbf{z}^{fin.}$ related by $\mathbf{z}^{fin.} = \phi \circ \mathbf{z}^{in.}$. In this context, neglecting collisions, the Vlasov equation is equivalent to the statement that
\begin{equation}\label{eqn:vlasov_flow}
f(\mathbf{z}^{in.}) = f(\mathbf{z}^{fin.})
\end{equation}
for the phase space density $f$.

Particle pushing algorithms approximate the flow for a short time, the time step $h$, using a discretization of the equations of motion. This could be by means of Runge-Kutta methods, the Boris~\cite{boris:70} or Boris-related~\cite{vay:08} particle pushers, or with symplectic integrators for single-particle~\cite{chacon-golcher_neri:2008,forest:06,qin_guan:08,ruth:83} or self-consistent Hamiltonian systems~\cite{abell_etal:2017,evstatiev_shadwick:2013,lewis_sykes_wesson:72,lewis:70a,lewis:70b,shadwick:14a,webb:16b}. It is thus well-known how to compute $\mathbf{z}^{fin.}$ from $\mathbf{z}^{in.}$ given the fields the particles move in. We therefore need only to determine the final distribution in eqn.~(\ref{eqn:vlasov_flow}) in some computable representation.

%% file: sections/algorithm.tex


The central concept of this algorithm is to take the relationship in eqn.~(\ref{eqn:vlasov_flow}) and develop a discrete representation using local basis functions. There are many options for representing the phase space density function, from Fourier series to splines. The derivation that follows applies to an arbitrary choice of basis functions, although we will in the examples specialize to localized tent functions instead for the sake of concreteness. 

We discretize the phase space density as a set of overlapping basis functions:
\begin{equation}
f(x, p) = \sum_j w_j \Lambda(x - x_j) \Lambda(p - p_j).
\end{equation}
from eqn.~(\ref{eqn:vlasov_flow})
\begin{equation}
f(x_{in.}, p_{in.}) = f(x_{fin.}, p_{fin.})
\end{equation}
Under this scheme, the individual $(x_j, p_j)$ coordinates change by $(\delta x_j, \delta p_j)$ in a single time step. We can then write eqn.~(\ref{eqn:vlasov_flow}) in terms of the local basis functions
\begin{equation}
\sum_j w_j^{(fin.)} \Lambda(x - x_j) \Lambda(p - p_j) = \sum_k w_k^{(in.)} \Lambda(x - x_k - \delta x_k) \Lambda(p - p_k - \delta p_k).
\end{equation}
By premultiplying by $\Lambda (x - x_{j'}) \Lambda(p - p_{j'})$, we can reduce this to a matrix equation for a potentially overcomplete basis set:
\begin{equation}
\sum_j \boldsymbol{\Lambda}_{j j'} w_j^{(fin.)} = \sum_k \boldsymbol{\Delta}_{k j'} w_k^{(in.)}
\end{equation}
where
\begin{equation}
\boldsymbol{\Lambda}_{j j'} = \int dx dp ~ \Lambda(x - x_j) \Lambda(p - p_j) \Lambda(x - x_{j'}) \Lambda(p - p_{j'})
\end{equation}
and
\begin{equation}
\boldsymbol{\Delta}_{k j'} = \int dx dp ~ \Lambda(x - x_k - \delta x_k) \Lambda(p - p_k - \delta p_k) \Lambda(x - x_{j'}) \Lambda(p - p_{j'}).
\end{equation}
Note that if we choose an orthogonal basis set, then $\boldsymbol{\Lambda}$ will be diagonal. However, we are not required to make this choice. The update sequence for each $w_j^{(fin.)}$ is then given by the matrix equation
\begin{equation}
w_j^{(fin.)} = \sum_{j', k} \boldsymbol{\Lambda}_{j j'}^{-1} \boldsymbol{\Delta}_{k j'} w_k^{(in.)}
\end{equation}
that depends on the matrix inverse of $\boldsymbol{\Lambda}$ -- which, will also be diagonal if we use an orthogonal basis -- and $\boldsymbol{\Delta}$, which contains the characteristic trajectory information over the time step. 

Global phase space volume conservation comes from the extent to which 
\begin{equation}
\sum_j w_j^{(fin.)} =^? \sum_j w_j^{(in.)}.
\end{equation}
The global conservation can therefore be written with the condition
\begin{equation}
\sum_{j} \left (\sum_k \boldsymbol{\Gamma}_{j, k} w_k^{(in.)} \right ) = \sum_j w_j^{(in.)}
\end{equation}
where $ \boldsymbol{\Gamma}_{j, k} = \sum_{j'}\boldsymbol{\Lambda}_{j j'}^{-1} \boldsymbol{\Delta}_{k j'}$. This is not satisfied for arbitrary choice of local basis functions, so the algorithm will not in general conserve the total phase space volume. The matrix $\boldsymbol{\Gamma}$ also has implications for positivity, and indeed there is no guarantee that the weights will remain positive for an arbitrary choice of basis. We will see an instance in our examples where negativity in the weights appears at a discontinuity in the phase space density.

The primary computational demand is in computing the matrix $\sum_{j'} \boldsymbol{\Lambda}_{j j'}^{-1} \boldsymbol{\Delta}_{k j'}$, as both are $N \times N$-dimensional matrices for $N$ unique phase space volume cells. $\boldsymbol{\Delta}$ must be recomputed every time step, while $\boldsymbol{\Lambda}^{-1}$ is a constant throughout the simulation. The convolving integrals that define $\boldsymbol{\Lambda}$ and $\boldsymbol{\Delta}$ integrate over smooth functions the local phase space density, so that filamentation on scales smaller than the grid resolution are smoothed across the grid and do not present a computational challenge.

Because this approach is built on established particle pushers, it can be extended to arbitrary dimensionality. The only dimension-dependent computations are the $\boldsymbol{\Lambda}$ and $\boldsymbol{\Delta}$ matrices, and those require integrating local regular basis shapes -- these integrals are straightforward to compute analytically. Furthermore, because the approach requires only a high-dimensional grid and is otherwise built on particle-in-cell techniques, the approach could be well-suited as an addition to mature particle-in-cell codes.

%% file: sections/examples.tex


We demonstrate this algorithm on two example problems -- the harmonic oscillator and pendulum problems, and the one-dimensional FEL equations. In each simulation we assume periodic boundaries in the coordinates and open boundaries in the momenta. As a metric for performance and convergence, we will look at global phase space volume conservation and entropy. We expect to see some loss of phase space volume, as volume elements that pass outside the range of the momentum grid will be lost.

We do not make any comments on the timing performance of the algorithm, as the implementation is more for proof-of-concept than to benchmark performance or, for example, scalability on many-core architectures. We do not that the basis matrix computations can be quite expensive -- $\boldsymbol{\Lambda}^{-1}$ and $\boldsymbol{\Delta}$ are both $MN \times MN$ matrices, $\boldsymbol{\Delta}$ is a sparse matrix that must be recalculated every step, and $\boldsymbol{\Lambda}^{-1}$ is the inverse of a sparse matrix, which therefore may not be itself sparse although it may be pre-computed once at the beginning of the simulation, although storing $8 \times (MN)^2$ double-precision numbers in memory may be prohibitive for large grids.


%% file: sections/pendulum.tex


\subsection{Harmonic Oscillators}

\begin{figure}
\centering
\includegraphics[width=\columnwidth]{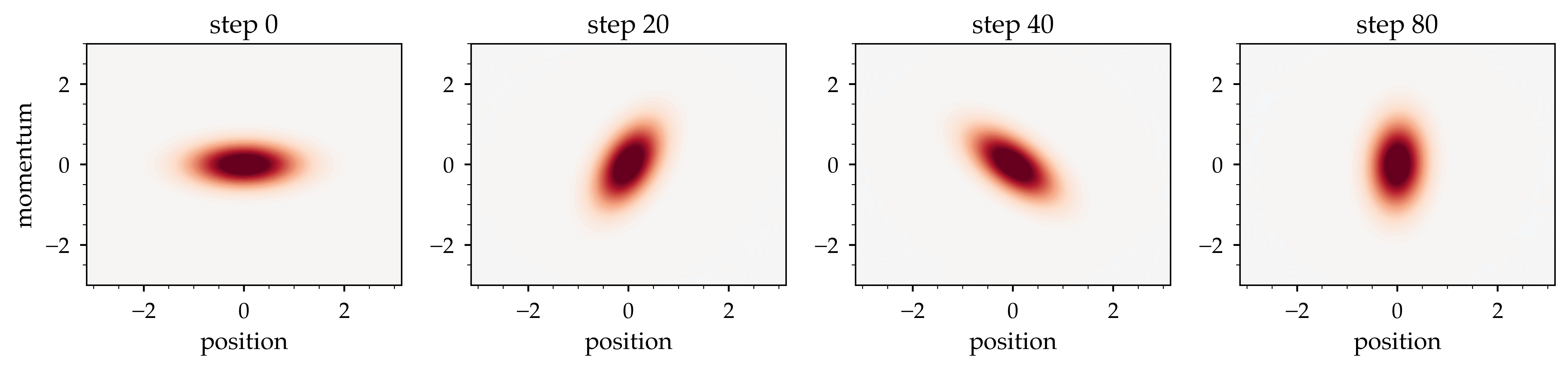} 
\caption{Phase space evolution of a linear harmonic oscillator.}
\label{fig:vlasov_sho}
\end{figure}

In order to test that the algorithm is properly handling the phase-space deposition we tested both a linear harmonic oscillator and a nonlinear harmonic oscillator. For the linear oscillator we analytically advanced the phase space distribution to understand any inherent diffusion in the algorithm. Figure~\ref{fig:vlasov_sho} shows the evolution of a gaussian initial phase space through one revolution around the origin. In this case the phase advance was  $\pi/10$. 
 
\begin{figure}
\centering
\includegraphics[width=\columnwidth]{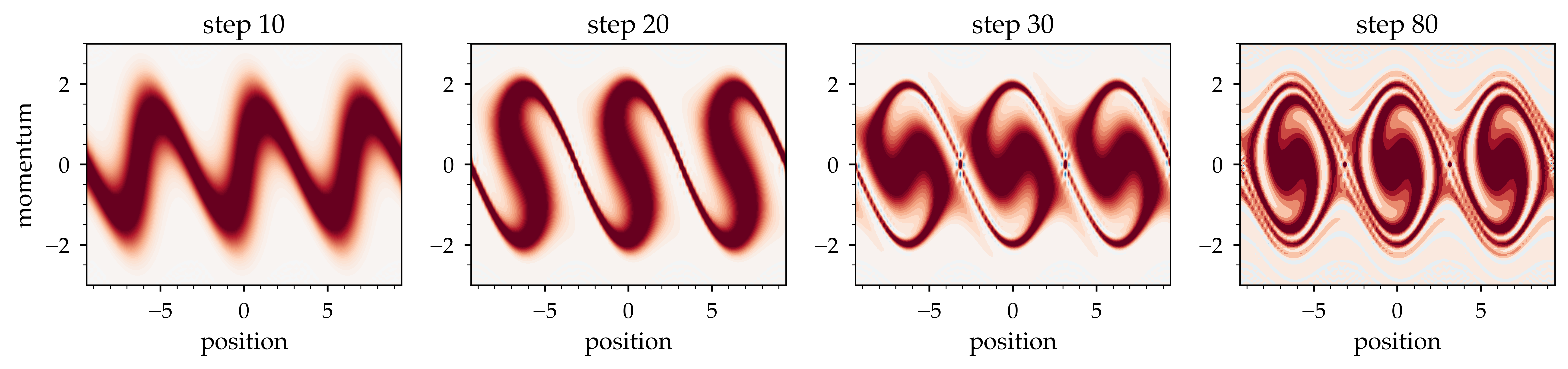} 
\caption{Phase space evolution of the pendulum equation with a second order symplectic integrator.}
\label{fig:vlasov_bucket}
\end{figure}

Our second benchmark applied a symplectic integrator to a nonlinear oscillator. Here we applied a second order integrator to the pendulum equation to demonstrate filamentation in phase space over long time scales.  The Hamiltonian for our problem was $\mathcal{H} = p^2/2 + \cos(x)$. Figure~\ref{fig:vlasov_bucket} shows the initial phase space and the phase space at four snapshots along the simulation. The integration step was $2 \pi / 20$ and the snapshots were taken at steps 10, 20, 30, and 80 going from left to right.

The colormap for this figure is red for positive values and blue for negative values. In the pendulum example, negative weights arise and propagate from regions where filamentation is quite strong. We can see this effect occurring next to the hyperbolic fixed points in the plot of step 30 in fig.~(\ref{fig:vlasov_bucket}), which will have a sharp discontinuity in the phase space density as any initial density beginning there will not move, while all the neighboring density values will tend to dilute. 

\begin{figure}
\centering
\includegraphics[width=0.5\columnwidth]{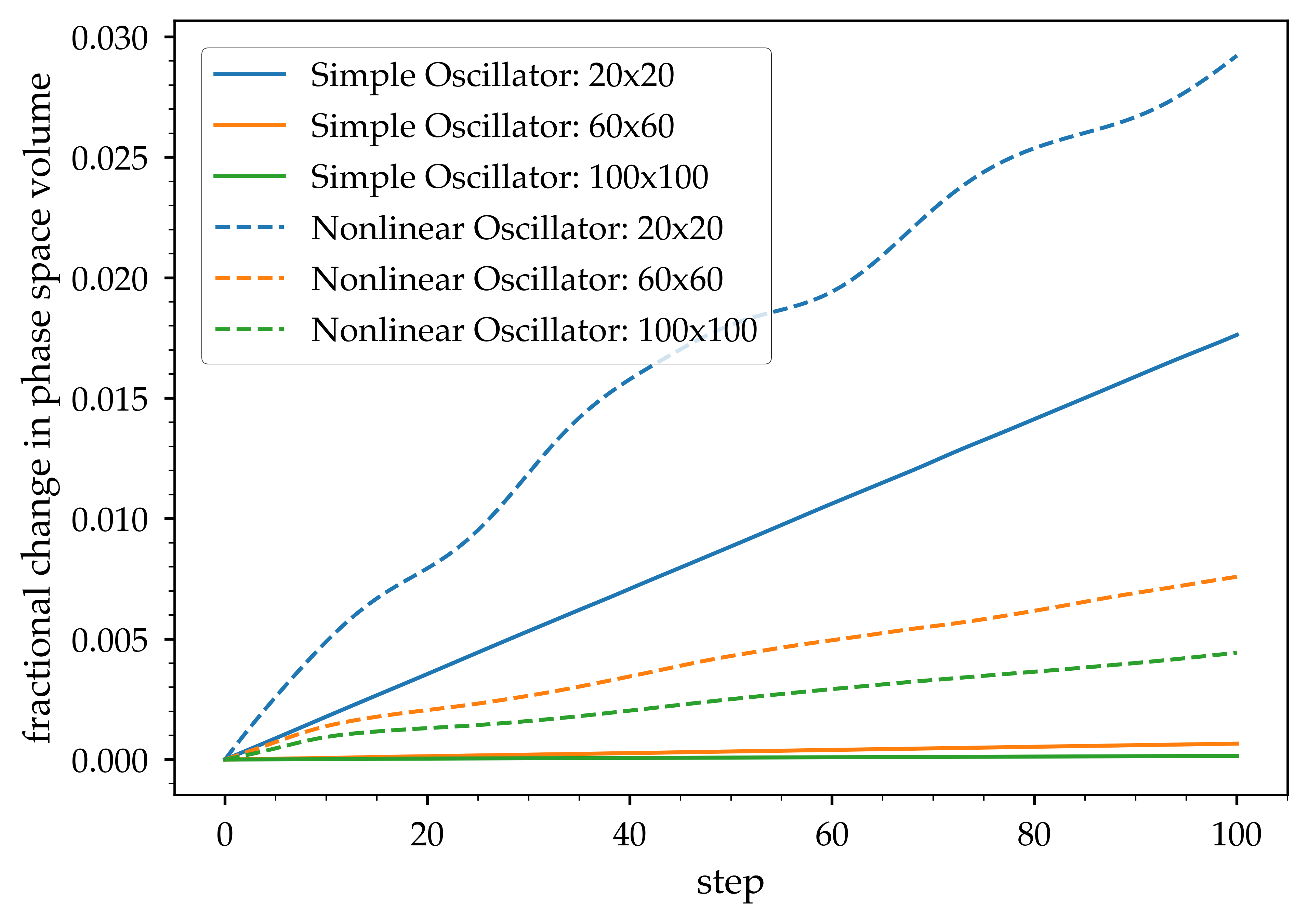} 
\caption{Fractional change in the sum of the weights as a function of time for different grid resolutions for both the linear and nonlinear oscillator.}
\label{fig:psp_area}
\end{figure}

In these regions, the grid is under-resolving a phase space feature, creating local oscillatory values in the phase space density. Figure~\ref{fig:psp_area} shows fractional change in the sum of the weights as a function of time for both the linear and nonlinear oscillator. Remarkably, the \emph{total} volume of phase space is well-conserved, even with the introduction of negative weights. The oscillatory behavior is similar in character to the Gibbs phenomenon in the Fourier analysis of discontinuous functions, but which can also be seen in, for example, electrostatic Poisson solvers on distributions with hard edges. 

\begin{figure}
\centering
\includegraphics[width=0.5\columnwidth]{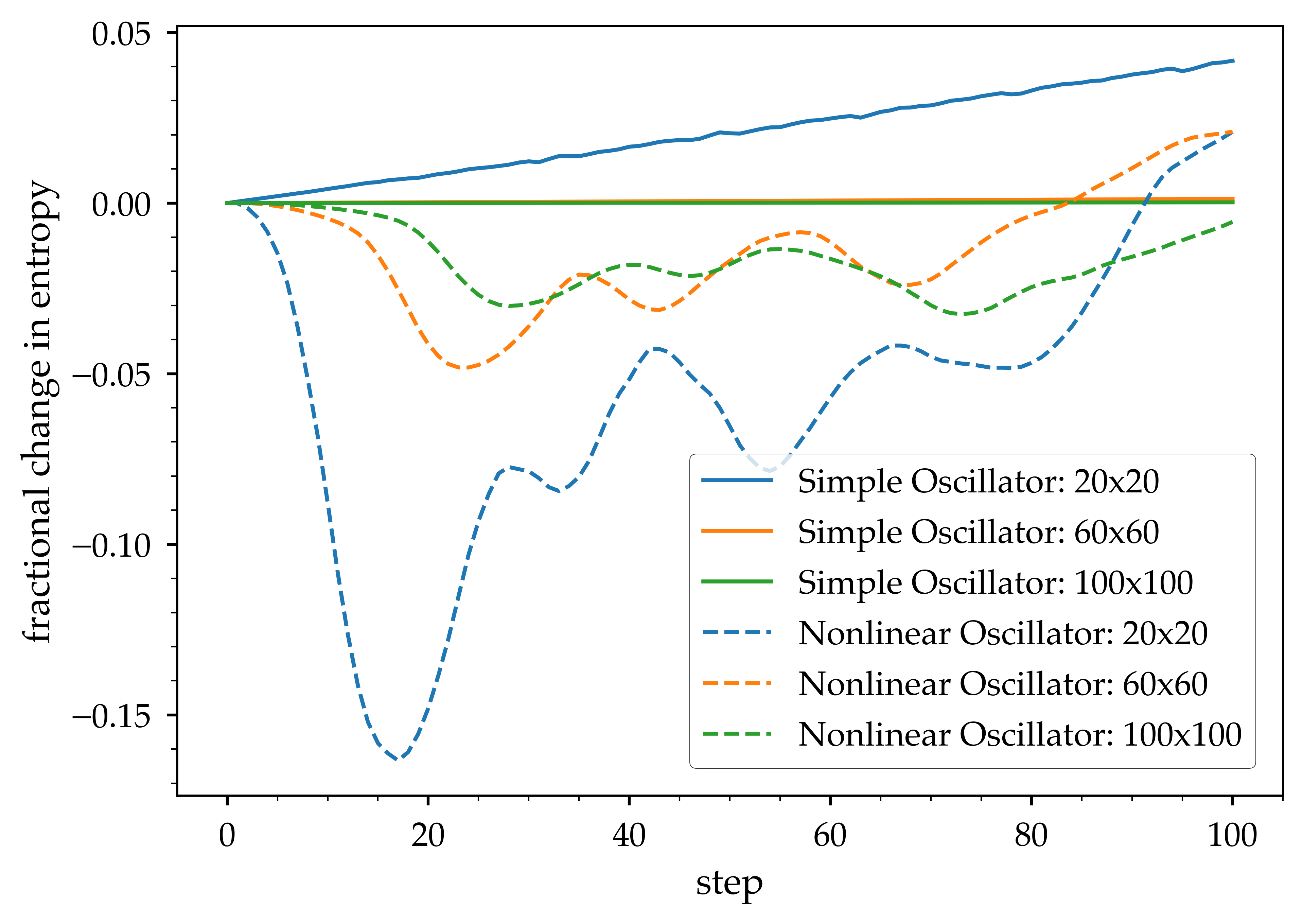} 
\caption{Fractional change in the system entropy as a function of time for different grid resolutions for both the linear and nonlinear oscillator.}
\label{fig:entropy}
\end{figure}

These discontinuities in phase space at the fixed point can be well described by an increasing entropy of the system. Figure~\ref{fig:entropy} shows the fractional change in entropy for both the linear and nonlinear oscillators. For the linear oscillator as expected there is no change in the system's entropy while for the nonlinear oscillator we are able to model the increase in entropy due to these hard edges in the phase space evolution.

%% file: sections/fel.tex


\subsection{FEL simulations} 

Next we simulated a FEL by numerically integrating the 1-D normalized FEL equations using a fourth-order Runge-Kutta algorithm. 

\begin{subequations}
\begin{equation}
{d\Phi \over dz } = \delta + \eta(z)
\end{equation}
\begin{equation}
{d\eta \over dz} = - \left[ \left( A(z)  + i\sigma^2 \langle e^{-i\Phi(z)} \rangle \right) e^{i \Phi(z)}  + c.c.\right] 
\end{equation}
\begin{equation}
{d A \over dz} = \langle e^{-i\Phi(z)} \rangle 
\end{equation}
\end{subequations}

\begin{figure}
\centering
\includegraphics[width=\columnwidth]{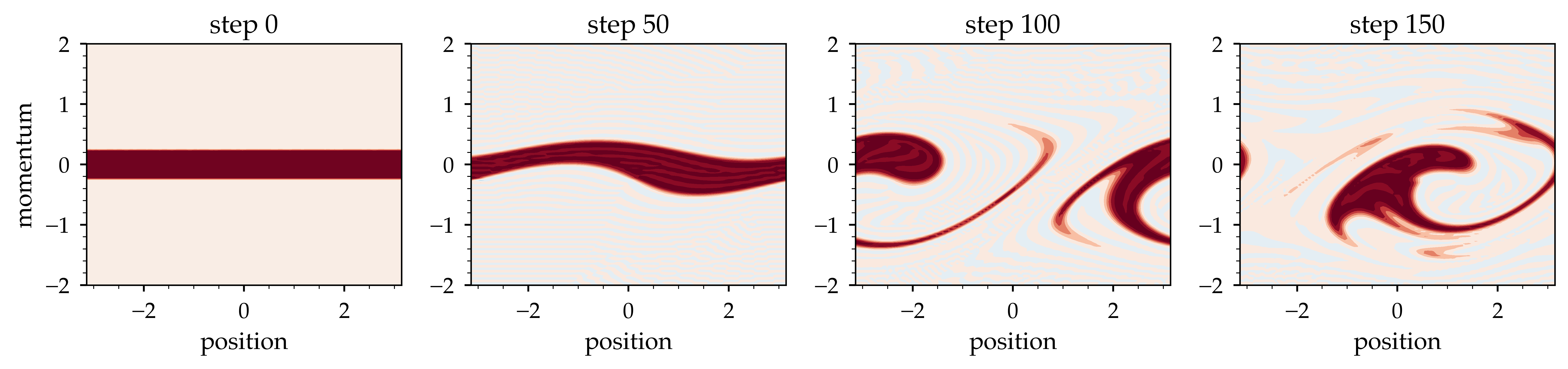} 
\caption{Phase space evolution for a self-consistent FEL, flat top initial distribution.}
\label{fig:fel_flattop}
\end{figure}

\begin{figure}
\centering
\includegraphics[width=\columnwidth]{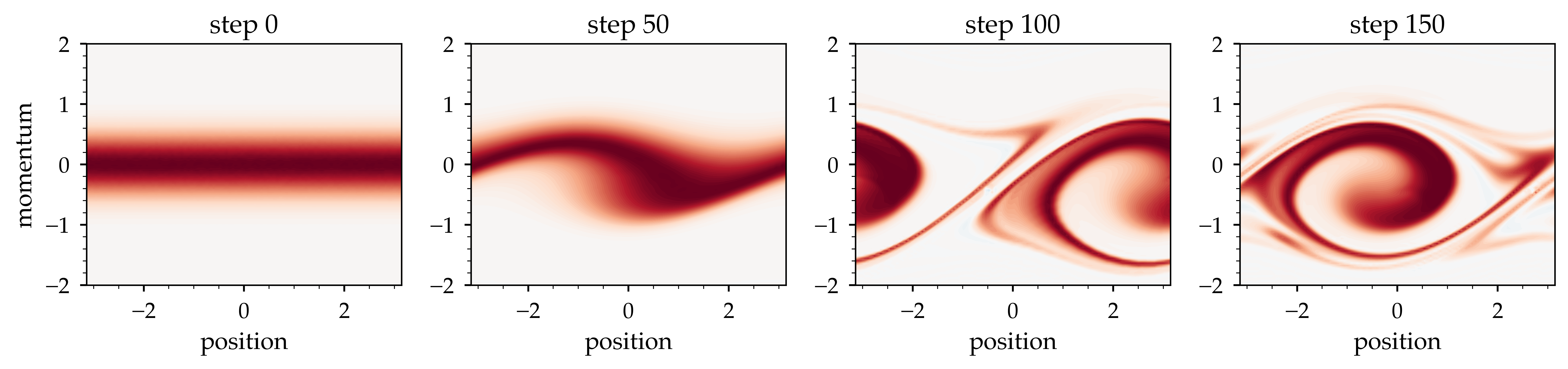} 
\caption{Phase space evolution for a self-consistent FEL, gaussian initial distribution.}
\label{fig:fel_gaussian}
\end{figure}

Here $\Phi$ is the ponderomotive phase which is the position coordinate in phase space, $\eta$ is the normalized energy, $A$ is the self-consistent field magnitude, and $z$ is our independent variable. Using these equations and our Vlasov code we simulated a FEL starting up from numerical noise. To examine the effect of hard edge distributions and negative weighs we simulated to distributions. One with a hard edge in energy, Figure \ref{fig:fel_flattop}, and one with a Gaussian distribution in energy \ref{fig:fel_gaussian}. Figures  \ref{fig:fel_flattop} and \ref{fig:fel_gaussian} show the initial phase-space distribution and the phase-space after 50, 100, and 150 steps. After 50 steps we can see the beginnings of phase-space bunching which becomes more prominent after 100 steps. after 150 steps there is clear filamentation in the phase-space caused by saturation of the FEL process. 

\begin{figure}
\centering
\includegraphics[width=0.5\columnwidth]{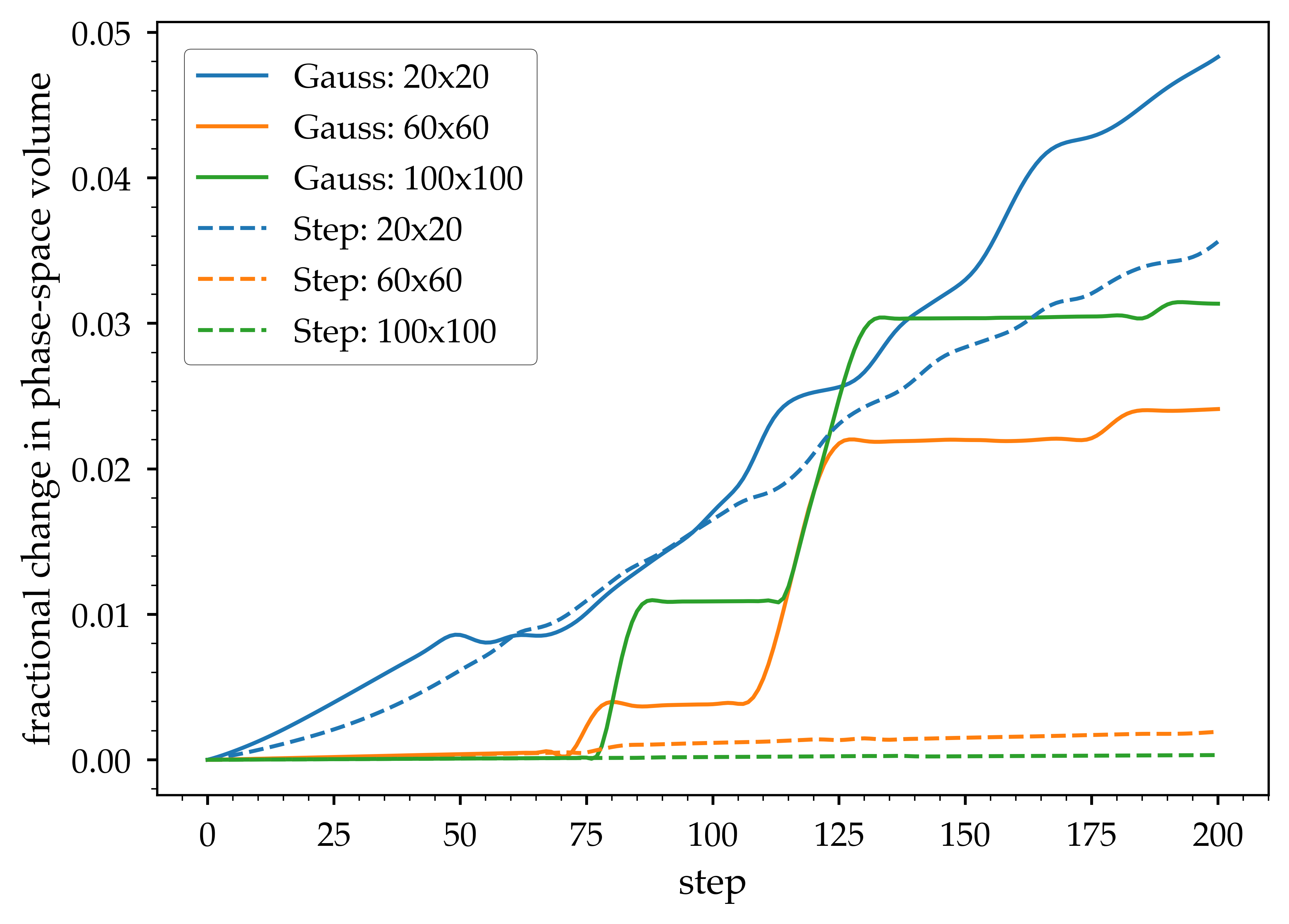} 
\caption{Fractional change in the phase-space density as a function of time step for the two different FEL simulations as we increase the mesh density. }
\label{fig:fel_weights}
\end{figure}

We can also see the same oscillatory behavior clearly in the evolution of a distribution that has a hard edge in fig.~(\ref{fig:fel_flattop}), where the distribution is initially uniform for momentum between $\pm .1$ and zero everywhere else. The oscillatory weights appear almost immediately. In contrast, with a gaussian initial distribution, fig.~(\ref{fig:fel_gaussian}), the oscillatory behavior and negative weights are almost completely absent. Figure \ref{fig:fel_weights} shows the fractional change in the sum of the weights for the two FEL problems as a function of integration step. Here for both distribution types the phase space density is well conserved as we increase the grid density. The Gaussian has discrete locations where the phase-space density changes, this is due to loss at the boundary in energy as these are open boundaries.

\begin{figure}
\centering
\includegraphics[width=0.5\columnwidth]{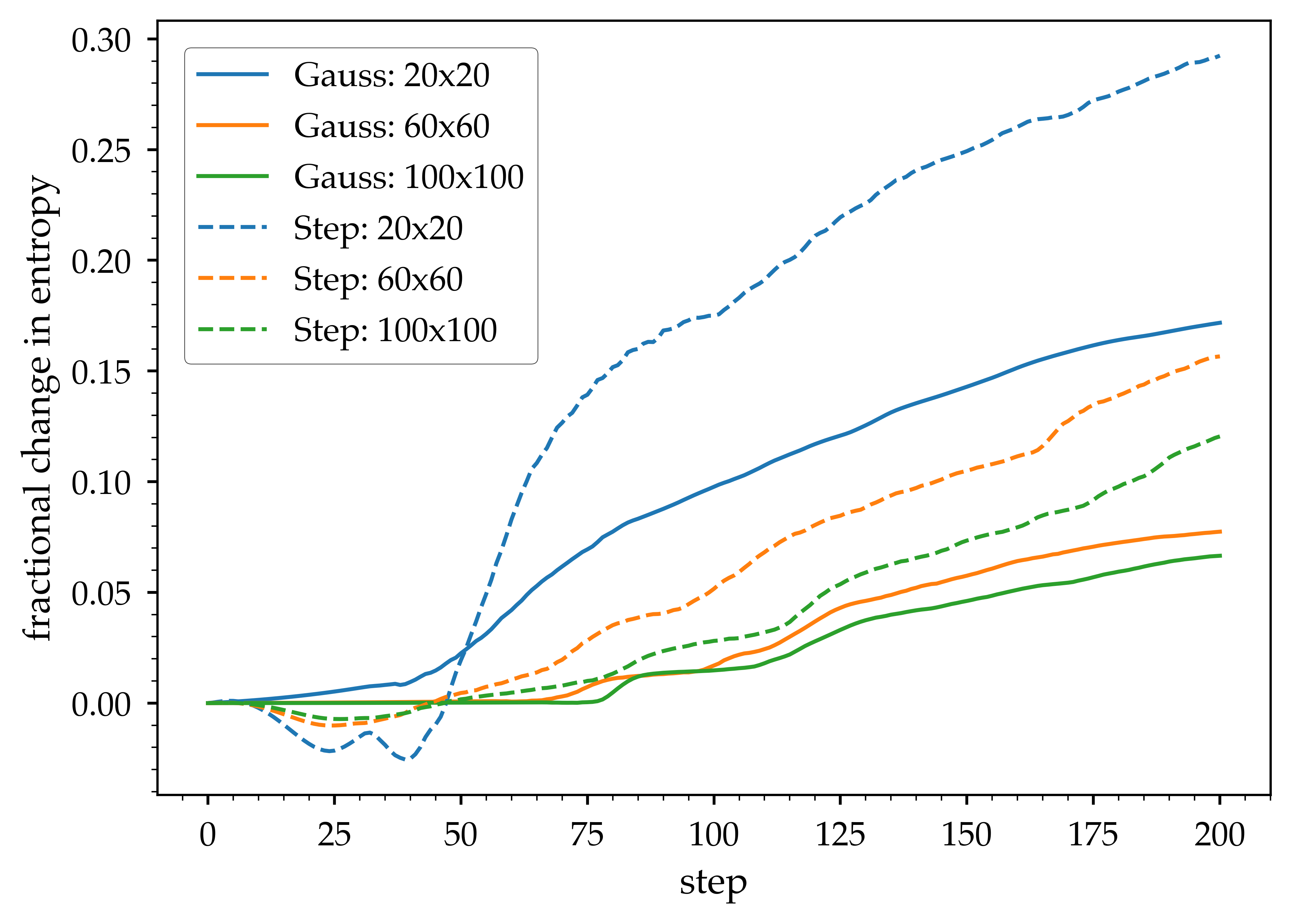} 
\caption{Fractional change in the entropy as a function of time step for the two different FEL simulations as we increase the mesh density.}
\label{fig:fel_ent}
\end{figure}

Figure \ref{fig:fel_ent} shows the fractional change in entropy as a function of time as we increase the mesh density. For the Gaussian distribution the entropy does not change significantly for high grid density until we begin seeing a change in phase-space density due to open boundaries in energy. However for the flat-top distribution we see a conserved phase-space density while the entropy oscillates. This is attributed to the negative weights introduced by evolving a top-hat distribution on a finite grid. 

\begin{figure}
\centering
\includegraphics[width=0.75\columnwidth]{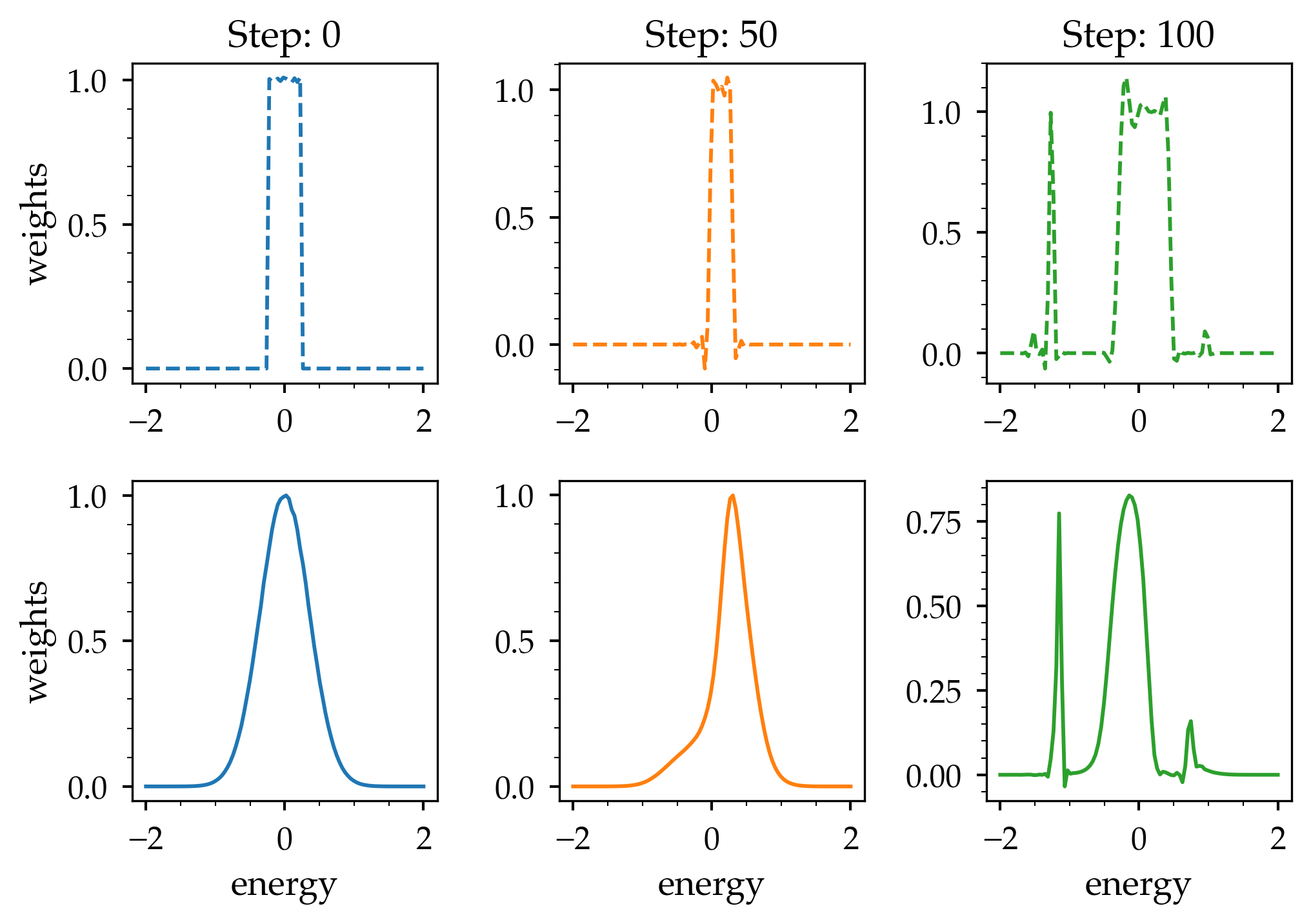} 
\caption{Line-out plot of the weights from the FEL simulations at three different time-steps. Top is the for the flat-top distribution in energy and bottom is for the Gaussian distribution in energy.}
\label{fig:fel_lineout}
\end{figure}

This Gibbs-like phenomenon resulting by depositing a hard edged distribution onto a finite grid is further demonstrated by examining a line out of the weights as a function of energy. Figure \ref{fig:fel_lineout} shows the weights as a function of energy for a fixed position for three different time steps. The negative weights are present shortly after the distribution starts evolving for the hard edged distribution while they only become apparent for the Gaussian distribution when strong filamentation occurs in the simulations. 

%% file: sections/discussion.tex


We have presented a new approach to deriving algorithms for solving the Vlasov equation on a gridded representation of phase space, based on tracking the flows of the phase space grid points and projecting the shifted local basis functions back onto the original basis. This approach uses particle tracking algorithms from particle-in-cell or other techniques to solve what is conventionally treated as a fluid equation.

This approach is explicit in time, with the initial and final weights of a time step related by a matrix equation. Those matrices are determined by convolution integrals with basis functions for the phase space density, which are completely general in principle. This flexibility could allow for clever choices in basis functions for specific applications that capture essential physics intrinsically, allowing for a Vlasov algorithm that is particularly efficient for certain types of problems.

We demonstrated this algorithm in a handful of test cases -- the linear oscillator and pendulum equations where the trajectories are single-particle, and for the one-dimensional FEL equations, a plasma instability. For our implementation, we chose to use local tent functions. This specific implementation showed oscillatory behavior and negative-valued weights emerging from under-resolved phase space features, such as discontinuities in the local density or phase space filaments that are on the order of a grid size. This feature was not present when the distribution is smooth. It is possible that smoother basis functions will ameliorate this effect.

A potential extension of this scheme is to provide a formal way to bridge the gap between moment-based fluid equations and full Vlasov equations. For example, if the phase space density is assumed to be locally thermal, so that the distribution is given by
\begin{equation}
f(v, x) = \frac{N}{\sqrt{2 \pi \sigma(x)^2}} \exp \left \{ - \frac{\left (v- \overline{v}(x) \right )^2}{2 \sigma(x)^2} \right \}
\end{equation}
then we can formally relate changes in $\sigma$ to changes in $v^2$ by carefully selecting local basis functions
\begin{equation}
f = \sum_j \frac{w_j}{\sqrt{2 \pi \sigma(x_j)^2}} \exp \left \{ - \frac{\left (v- \overline{v}(x_j) \right )^2}{2 \sigma(x_j)^2} \right \}.
\end{equation}
This may provide a formal method to stitch together various high-order fluid models across a simulation domain based on the local physics.